\documentclass[prb,aps,amsmath,twocolumn,amssymb,floatfixng,showpacs,
superscriptaddress,footinbib]{revtex4-1}
\pdfoutput=1
\usepackage[dvips]{graphics}
\usepackage{bm,amsfonts}
\usepackage{enumerate}
\usepackage{color}
\usepackage{graphicx}
\usepackage[colorlinks=true,linktoc=page,linkcolor=blue,citecolor=magenta]{hyperref}

\newcommand\bea{\begin{eqnarray}}
\newcommand\eea{\end{eqnarray}}
\newcommand\beq{\begin{equation}}  
\newcommand\eeq{\end{equation}}
\newcommand{\non}{\nonumber} 
\newcommand{\noin}{\noindent} 
\newcommand{\ie}{{\it i.e.$~$}}
\newcommand{\eg}{{\it e.g.}}
\newcommand{\etal}{{\it et al.}}

\begin{document}

\title{Transport and noise properties of a normal metal$-$superconductor$-$normal metal junction
with mixed singlet and chiral triplet pairings}

\author{Ganesh C. Paul}
\email{ganeshpaul@iopb.res.in}
\author{Paramita Dutta}
\email{paramitad@iopb.res.in}
\thanks{First two authors have contributed equally to this work.}
\author{Arijit Saha}
\email{arijit@iopb.res.in}

\affiliation{Institute of Physics, Sachivalaya Marg, Bhubaneswar-751005, India} 

\begin{abstract}
We study transport and zero frequency shot noise properties of a normal metal-superconductor-normal metal (NSN) junction, 
with the superconductor having mixed singlet and chiral triplet pairings. We show that in the subgapped regime
when the chiral triplet pairing amplitude dominates over that of the singlet, a resonance phenomena emerges out
at zero energy where all the quantum mechanical scattering probabilities acquire a value of 0.25. 
At the resonance, crossed Andreev reflection mediating through such junction, acquires a zero energy peak. 
This reflects as a zero energy peak in the conductance as well depending on the doping concentration. 
We also investigate shot noise for this system and show that shot noise cross-correlation is negative in the subgapped 
regime when the triplet pairing dominates over the singlet one. The latter is in sharp contrast to the positive shot noise 
obtained when the singlet pairing is the dominating one. 
\end{abstract}


\maketitle

\section{Introduction} 

Study of transport signatures at the interface of normal metal-superconductor (NS) hybrid structures has been 
the subject of intense research interest during the last few decades~\cite{blonder1982transition,klapwijk2004proximity,ASahaReview}. 
The key issue behind the low-energy quantum transport phenomena in this type of hybrid junction is the process of 
Andreev reflection~\cite{andreev1964thermal}. When a normal metal electron with energy below the superconducting 
gap regime incident on the NS interface, a hole with opposite spin reflects back from the interface and 
as a result a Cooper pair jumps into the superconductor. Such reflection process is called the phenomenon of 
Andreev reflection (AR)~\cite{andreev1964thermal} in literature. 
 
Another intriguing phenomenon occurs in case of a normal metal-superconductor material-normal metal (NSN)
junction in which an electron incident from one of the normal metal leads forms a pair with another electron 
from the other normal metal lead and jumps into the superconductor as a Cooper pair. Such non-local process
is called crossed Andreev reflection (CAR)~\cite{falci2001correlated,bignon2004current,recher,belzig,yeyati,das2008spintronics} 
whose signature has been verified in various experiments~\cite{russo2005experimental,chandrasekhar,WeiChandrasekhar,AndyDas,LHofstetter,
LGHerrmann,ZimanskyChandrasekhar,JBrauer}. CAR can also be used to produce non local entangled electron pairs~\cite{recher,LesovikMatin,yeyati}.
From the practical point of view, superconducting hybrid structures can be designed 
by placing a bulk superconducting material in close proximity to a normal metal system~\cite{russo2005experimental,chandrasekhar,
WeiChandrasekhar,ZimanskyChandrasekhar,JBrauer} and superconducting correlation is actually induced into the non-superconducting 
region via the proximity effect. 

Till date, most of the research works have been carried out using either conventional spin-singlet~\cite{falci2001correlated,ASahaReview} 
or spin-triplet~\cite{yamashiro1998theory,honerkamp1998andreev,sengupta2006spin} superconductor with the orbital 
even and odd parity, respectively. Nevertheless this classification of the superconductors holds as long as they maintain 
inversion symmetry~\cite{yip2014noncentrosymmetric}. A different situation arises when we consider an 
unconventional superconductor without inversion symmetry. The physical properties of such superconductors with broken inversion 
symmetry becomes interesting due to the mixing of spin-singlet and spin-triplet order parameter without any 
parity symmetry. Non-centrosymmetric superconductors~\cite{bauer2012non,YanaseSigrist} (NCS) are examples of such superconductors 
where the spin-singlet and triplet pairing mixing~\cite{nishiyama2007spin} is present with time reversal symmetry but with 
broken inversion symmetry~\cite{sigrist1991phenomenological,tanaka2011symmetry,frigeri2004superconductivity}. The absence of the 
parity symmetry in NCS may lead to several interesting properties determined by the ratio of the amplitudes of the 
spin-singlet to spin-triplet pair potentials~\cite{burset2014transport}. Among such properties emergence of topological
spin current in NCS superconductor~\cite{Nagaosa1,LuYip}, magneto-electric effects~\cite{yip2014noncentrosymmetric}, 
magnetism~\cite{YanaseSigrist} etc. have been reported in recent times. 
These interesting properties have drawn attention of the community towards exploring the nature of NCS superconductor
and as a consequence, the list of NCS materials is growing gradually~\cite{karki2010structure}. 
Another reason behind this attraction is that recently, it has been shown that this type of superconductor 
characterized by time-reversal symmetry can hold an even number of Majorana 
Fermions~\cite{sato2009topological,santos2010superconductivity,duckheim2011andreev}. Therefore, further 
investigations are required to explore the properties of NCS as well as the effect of NCS on transport phenomena 
through superconducting hybrid junctions.

Very recently transport signature of NS and superconductor-normal metal-superconductor (SNS) junction with mixed singlet and
chiral triplet pairing has been reported by Burset \etal~\cite{burset2014transport}. They obtain a zero-energy peak in
the conductance in a NS junction when the triplet pairing is the dominating one over the singlet part. However, NSN junction 
and the properties of CAR in the above context has never been studied so far. The latter motivated us to investigate transport 
and shot noise phenomena through a NSN junction in which a one-dimensional (1D) nanowire (NW) is placed in close proximity to a 
superconductor which contains a pair potential of mixed singlet and chiral triplet type. The NW is attached to two 
normal metal (N) leads. We incorporate three regimes corresponding to the amplitude of the spin-singlet part being lesser, 
equal to and larger than that of the spin-triplet part. We adopt Blonder-Tinkham-Klapwijk (BTK) formalism~\cite{blonder1982transition} 
to calculate the quantum mechanical scattering amplitudes through the junction and conductance, shot noise therein. 
We find zero-energy peak in the conductance depending on the degree of mixing of the pair potentials and the doping. 
We also calculate the zero frequency shot noise (auto and cross-correlation) and show that the shot noise cross-correlation 
becomes positive to negative as long as the triplet pairing dominates over the singlet one. 

The remainder of this paper is organized as follows. In Sec.~\ref{sec:II} we describe our model. 
Sec.~\ref{sec:III} is devoted to the scattering matrix (BTK) formalism by which we calculate the quantum mechanical 
scattering amplitudes to obtain conductance and shot noise through the NSN junction.
We present our numerical results in Sec.~\ref{sec:IV} which includes scattering probabilities, conductance and 
shot noise for different parameter regimes. 
Finally, we summarize and conclude in Sec.~\ref{sec:V}.

\section{Model} 
\begin{figure}[!thpb]
\centering
\includegraphics[width=0.95\linewidth]{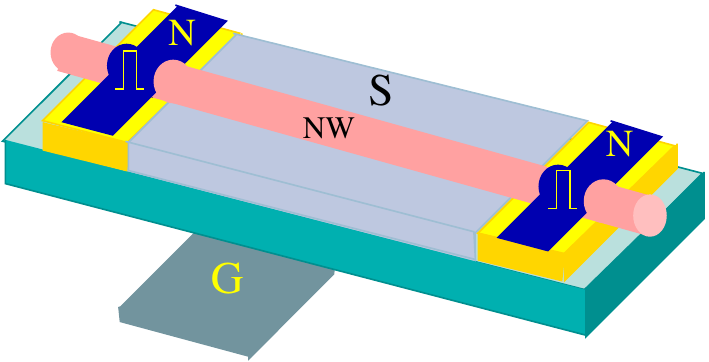}
\caption{(Color online) Schematic of the quasi one-dimensional NSN setup in which a nanowire (NW) (pink, light grey) is placed 
in close proximity to a bulk superconductor (light brown, light grey) and superconductivity is induced in the NW
via the proximity effect. The NW is attached to two normal (N) metal leads (blue, black). The gate G (light cyan, light grey) 
controls the chemical potential in the NW. Two $\delta$-function barriers are symbolically denoted by the two yellow (light grey) 
rectangular barriers at each N-NW interface.}
\label{model}
\end{figure}

{\label{sec:II}}
In Fig.~\ref{model} we present the schematic of our proposed set-up in which a 1D NW is placed in close proximity to
a bulk superconducting material. Here superconductivity is induced in the NW via the proximity effect. The NW is attached 
to two normal metal leads. A gate voltage $G$ can tune the chemical potential inside the NW. 
Instead of conventional s-wave superconductor here we consider the pairing potential
of the superconductor as a combination of spin-singlet and chiral spin-triplet states. 
We choose the $x$-axis along the direction of the NW. The two N-NW interfaces are located at $x=0$ 
and $x=L$ respectively. At each N-NW interface we consider an insulating barrier which is modeled 
by the $\delta$-function potential given as $V(x)=(\hbar^2k_F/m)Z\delta(x)$ where $k_F$ is the Fermi wave vector, 
$m$ denotes electron mass and $Z$ is the strength of the barrier. 

Hence the NSN junction can be described by the Bogoliubov-deGennes (BdG) equations as,
\bea
H(\mathbf{k}) \psi(\mathbf{k}) = \epsilon \psi(\mathbf{k})
\eea
where
\bea
H(\mathbf{k})=
\begin{bmatrix}
[E(\mathbf{k})-\mu)]\hat{\sigma}_0 & \hat{\Delta}(\mathbf{k})\\
\hat{\Delta}^\dagger(\mathbf{k}) & [\mu-E(-\mathbf{k})]\hat{\sigma}_0 
\end{bmatrix} \ .
\eea

Here $E(\mathbf{k})= (\hbar^2/2m)k_x^2+U$ is the dispersion relation of the electronic excitation measured from chemical potential 
$\mu$. $U$ is the electrostatic potential in the normal region. $\sigma_0$ is the $2 \times 2$ Identity matrix in spin space. 
We write the four component wave function in Nambu representation as 
$\psi(\mathbf{k})= [u_\uparrow(\mathbf{k}),u_\downarrow(\mathbf{k}),v_\uparrow(\mathbf{k}),v_\downarrow(\mathbf{k})]^T$ 
where $u_\sigma(\mathbf{k})$ and  $v_\sigma(\mathbf{k})$ are the electron and hole components respectively with spin 
$\sigma = \uparrow,\downarrow$ and $\mathbf{k}$ is the wave vector.

In the superconducting region, due to the presence of both spin-singlet and chiral spin-triplet states, the pairing potential 
$\hat{\Delta}(\mathbf{k})$ ($2 \times 2$ matrix) can be written in general 
$\hat{\Delta}(\mathbf{k})=i[\Delta_s(\mathbf{k})\hat{\sigma}_0 + 
\sum_{j=1}^{3}d_j(\mathbf{k})\hat{\sigma}_j]\hat{\sigma}_2 e^{i\phi}$. Here, $\hat{\sigma}_{1,2,3}$ are Pauli spin matrices operating 
on spin space and $\phi$ is the superconducting phase factor. The spin singlet pairing $\Delta_s(\mathbf{k})$ characterizes the 
conventional $s$ wave superconducting order parameter. Here we consider only the mean-field value of $\Delta_s(\mathbf{k})$ \ie
$\Delta_s(\mathbf{k})=\Delta_{s}$.
On the contrary, triplet pairing potential is described by an odd vector function as $\mathbf{d}(\mathbf{k})=-\mathbf{d}(-\mathbf{k})$.
 
Following Burset \etal~\cite{burset2014transport} we take chiral triplet state of the form,
\bea
\mathbf{d}(\mathbf{k})&=&\Delta_p\frac{k_x+i{\chi}k_y}{\lvert \mathbf{k}\rvert}
\hat{z} \non \\
&=&\Delta_{p}e^{i{\chi}\theta}\hat{z} \ ,
\eea 
where $\Delta_{p}$ is non-negative amplitude of the triplet pairing potential. $\chi$ determines the orientation of the angular momentum 
of the Cooper pairs and it can take $\pm$ sign corresponding to the parallel and anti-parallel direction respectively. $\theta$ 
represents the relative orientation between the singlet and chiral triplet pairing states. With this consideration paring potential now takes 
the form, $\hat{\Delta}(\mathbf{k})=i[\Delta_s\hat{\sigma}_0+\Delta_{p}e^{i{\chi}\theta}\hat{\sigma}_3]\hat{\sigma}_{2}e^{i\phi}$. 
This simple choice of the pairing potential takes into consideration the mixing of the spin-singlet and spin-triplet states. 
With this pairing potential, the band dispersion becomes~\cite{burset2014transport}
\bea
\epsilon_{1,2}(\mathbf{k})=\sqrt{E^2(\mathbf{k})+\Delta_s^2+\Delta_p^2\pm 2\Delta_s\Delta_p \cos{\theta}} \ ,
\label{delta}
\eea
which explicitly depends on the relative orientation of singlet and chiral triplet pairing components.
Now the full $4 \times 4$ Hamiltonian $H(\mathbf{k})$ can be written in block diagonal form which implies decoupling of two spin channels 
$\uparrow\downarrow$ and $\downarrow\uparrow$. Hence the effective pairing potentials corresponding to these two channels 
become~\cite{burset2014transport}
\bea
\Delta_{1,2}(\theta)=[\Delta_s \pm \Delta_{p}e^{i{\chi}\theta}]\hat{\sigma}_{2}e^{i\phi} \ .
\eea
Therefore, it will now be sufficient to consider two effective complex pair potentials $\Delta_{1,2}(\theta)$ among which 
$\Delta_2(\theta)$ vanishes for a particular choice of the $\Delta_s(=\Delta_p \cos \theta)$ and also it changes sign for $\Delta_s >\Delta_p$.

\begin{figure}[!thpb]
\centering
\includegraphics[width=0.85 \linewidth]{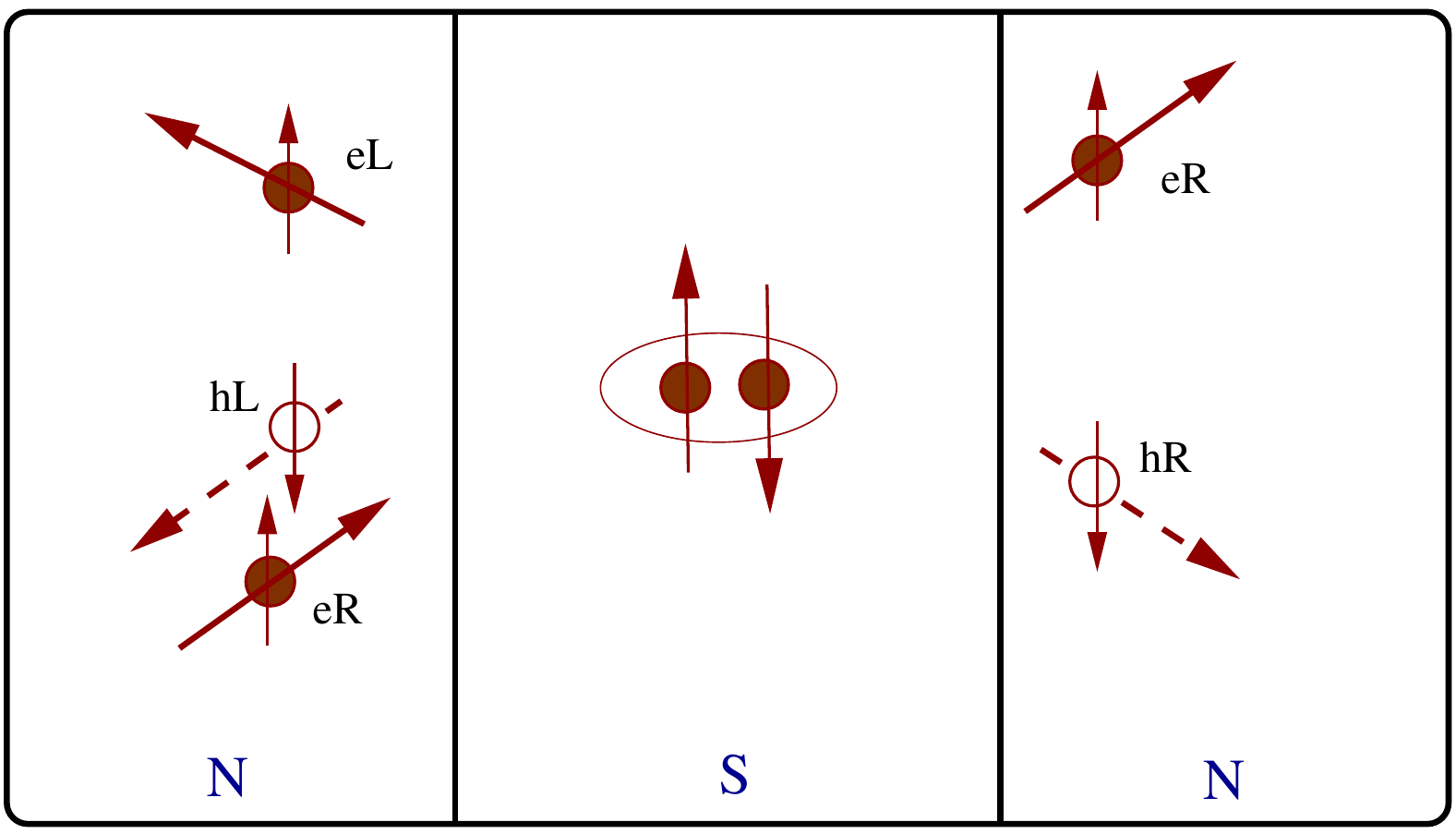}
\caption{(Color online) Schematic for the electron (solid sphere) and hole (hollow sphere) trajectories
(solid and dashed lines, respectively) corresponding to the four quantum mechanical scattering processes 
occurring at a NSN junction. Notations in the figure denote eR: right-moving electron; eL: left-moving electron; 
hR: right-moving hole; hL: left-moving hole.}
\label{nsn}
\end{figure}

In the quasi $1$D limit, electrons can propagate only in the $x$-direction with the transverse component
of the wave vector $k_y$ being conserved. Hence the band energy $E({\pm \mathbf{k}})$ can be written 
as $E(\pm k_x)$ for a particular choice of $k_y$. We choose $k_y=0$ for our analysis. Also, we assume that 
the band energies for the electrons moving to the left and right are equal to each other. We define right 
movers by $\theta^{+}=\theta$ and left movers by $\theta^{-}=\pi-\theta$. After decoupling for each spin channel, 
the BdG equations in the $2 \times 2$ form can be written as,
\begin{align}
\begin{bmatrix}
[E(\alpha \mathbf{k})-\mu] & s_{\sigma} \Delta_\sigma(\theta^\alpha)
e^{i\phi}\\
s_{\sigma} \Delta_\sigma^*(\theta^\alpha)e^{-i\phi} & 
[\mu-E(-\alpha \mathbf{k})]
\end{bmatrix}
\begin{bmatrix}
u_\sigma(\theta^\alpha)\\
v_\sigma(\theta^\alpha)
\end{bmatrix}
=\epsilon
\begin{bmatrix}
u_\sigma(\theta^\alpha)\\
v_\sigma(\theta^\alpha)
\end{bmatrix} 
\label{ham2}
\end{align}
where $\epsilon\geq 0$ is the excitation energy; $\alpha=\pm$ corresponds to the right and left 
movers; $s_{\sigma}=(-1)^{\sigma-1}$ and $\sigma=1,2$ denotes the different spin channels. Thus, 
the pairing potential is different for each independent spin channel as well as for direction of 
motion of particles. Also, the gap amplitude can be different depending on the direction of motion 
as argued in Ref.~\onlinecite{klapwijk2004proximity}.
Left mover with spin $\uparrow$ and $\downarrow$ will be effected by the pairing potential 
$\Delta_1(\theta^-)$ and $-\Delta_2(\theta^-)$ respectively. On the other hand, right mover will 
experience the effective pairing potential $\Delta_1(\theta^+)$ and $-\Delta_2(\theta^+)$ 
corresponding to $\uparrow$ and $\downarrow$ spin channels respectively. 

Electron and hole components of the wave functions are given by,
\bea
u_\sigma(\theta^\alpha)&=
\frac{1}{\sqrt{2}}\Big(1+\frac{\sqrt{\epsilon^2- {\lvert\Delta_\sigma(\theta^\alpha)\rvert}^2}}{\epsilon}\Big)\ ,\\
v_\sigma(\theta^\alpha)&=
\frac{1}{\sqrt{2}}\Big(1-\frac{\sqrt{\epsilon^2- {\lvert\Delta_\sigma(\theta^\alpha)\rvert}^2}}{\epsilon}\Big)\ .
\eea

\section{The Scattering Matrix} {\label{sec:III}}
In this section we present the scattering matrix obtained employing the BTK 
formalism~\cite{blonder1982transition} for our NSN geometry. Normal metallic region is described by 
considering $\hat{\Delta}(\mathbf{k})=0$ and also we set $U=0$ there to carry out our analysis. When an incident electron 
coming from one of the normal metal leads with energy below the superconducting gap scatters from the 
NS interface, the corresponding scattering phenomena can be described by four possible quantum mechanical 
processes. These processes are: (a) normal reflection of electron from the NS boundary 
(b) AR of incident electron as a hole in the same lead (c) elastic co-tunneling (CT) in which the incident electron 
transmits to the other lead as an electron and (d) transmission of hole in the other lead via the CAR process. 
The schematic of these processes are displayed in Fig~\ref{nsn}.

In order to obtain reflection, AR, CT and CAR amplitudes through the NSN junction we write the wave functions 
in the three regions as follow,

\bea
\psi_\sigma^L(x)=e^{i k_e x}
\begin{bmatrix}
1\\
0
\end{bmatrix}
+&r_\sigma(\epsilon)e^{-i k_e x}
\begin{bmatrix}
1\\
0
\end{bmatrix}\non\\
+&r_{h \sigma}(\epsilon)e^{i k_h x}
\begin{bmatrix}
0\\
1
\end{bmatrix},
\eea
\bea
\psi_\sigma^S(x)=&a_\sigma(\epsilon)e^{iq_ex}
\begin{bmatrix}
u_\sigma(\theta^+)e^{i\phi}\\
\eta_{\sigma}^*(\theta^+)v_\sigma(\theta^+)
\end{bmatrix} \non \\
+&b_\sigma(\epsilon)e^{-iq_ex}
\begin{bmatrix}
u_\sigma(\theta^-)e^{i\phi}\\
\eta_{\sigma}^*(\theta^-)v_\sigma(\theta^-)
\end{bmatrix}\non\\
+&c_\sigma(\epsilon)e^{-iq_hx}
\begin{bmatrix}
\eta_{\sigma}(\theta^+)v_\sigma(\theta^+)\\
u_\sigma(\theta^+)e^{-i\phi}\\
\end{bmatrix}\non\\
+&d_\sigma(\epsilon)e^{iq_hx}
\begin{bmatrix}
\eta_{\sigma}(\theta^-)v_\sigma(\theta^-)\\
u_\sigma(\theta^-)e^{-i\phi}\\
\end{bmatrix},
\eea
\bea
\psi_\sigma^R(x)=t_{\sigma}(\epsilon) e^{ik_ex}
\begin{bmatrix}
1\\
0
\end{bmatrix}
+
t_{h \sigma}(\epsilon) e^{-ik_hx}
\begin{bmatrix}
0\\
1
\end{bmatrix}
\eea
with $\eta_{\sigma}(\theta^\alpha)=s_\sigma\Delta_{\sigma}(\theta^\alpha)/
\lvert {\Delta_{\sigma}(\theta^\alpha)}\rvert$.  
Here $\psi_\sigma^L(x)$, $\psi_\sigma^R(x)$, $\psi_\sigma^S(x)$ are the 
wave functions for the left, right normal metal leads and the superconductor respectively. 
$k_e$ and $k_h$ are the wave vectors for the electron and hole respectively in the normal metal 
regions whereas, $q_e$ and $q_h$ are the same for the superconducting region. They can be expressed as,
\bea
&&k_{e(h)}= k_F \sqrt{(1\pm \epsilon/\mu)} \non \\
&&q_{e(h)}=k_F \sqrt{{(\mu+U) \pm i \sqrt{\lvert \Delta_{\sigma}\rvert^2-\epsilon^2}}\over{\mu}}
\label{keh}
\eea
where, $\Delta_{\sigma}$ can be $\Delta_1$ or $\Delta_2$ depending on the spin channel. 

Here, $r_{\sigma}$, $r_{h \sigma}$, $t_{e \sigma}$ and $t_{h \sigma}$ denote the normal reflection, AR, 
CT and CAR amplitudes respectively. They can be obtained by considering the boundary conditions for the wave 
functions such as, for the left boundary ($x=0$)
\bea
\psi^L\rvert_{x=0} &=& \psi^S\rvert_{x=0}~, \non \\
\partial_x\psi^S\rvert_{x=0}-\partial_x\psi^L\rvert_{x=0}&=&k_F Z\psi^L(0) 
~~~~~~ 
\eea
and for the right boundary ($x=L$),
\bea
\psi^R\rvert_{x=L} &=& \psi^S\rvert_{x=L}~, \non \\
\partial_x\psi^R\rvert_{x=L}-\partial_x\psi^S\rvert_{x=L} &=& k_F Z\psi^R(L).
~~~~~~ 
\eea
Numerically solving these eight equations we get the amplitudes corresponding to all scattering 
processes ($r_{\sigma}$, $r_{h \sigma}$, $t_{e \sigma}$ and $t_{h \sigma}$) and probability therein. 
We solve these  equations for each spin channel, $\sigma=1$ and $\sigma=2$. Here we denote 
$R_{e\sigma}=\lvert r_{\sigma} \rvert^2$, $R_{h\sigma}= {k_h \over k_e} \lvert r_{h \sigma} \rvert^2$,
$T_{e\sigma}=\lvert t_{\sigma} \rvert^2$ and 
$T_{h\sigma}={k_h \over k_e} \lvert t_{h \sigma} \rvert^2$ as the probability for normal reflection,
AR, CT and CAR respectively. All the probabilities together satisfy the unitarity relation,
\bea
T_{e\sigma}+T_{h\sigma}+R_{e\sigma}+R_{h\sigma}=1 \ .
\eea
for each spin channel ($\sigma$=1,2) separately. Note that, the factor $k_h/k_e$ is introduced 
in order to maintain the probability current conservation.

At zero temperature, conductance for a particular electron energy $\epsilon$ and a chiral angle 
$\theta$ can be found by taking contributions from both the spin channel $\sigma=1$ and $2$ using 
the following relation,
\bea
&&G(\epsilon,\theta)={e^2 \over h} \sum\limits_{\sigma} 
\left(\lvert t_{e \sigma}\rvert^2 -\lvert t_{h \sigma} \rvert^2 \right).
\eea
\noin
We normalize the conductance by the normal state conductance 
$G_0={2e^2 \over h} D(\theta) $, where, 
$D(\theta)= 4 \cos^2 \theta /(Z^2+4 \cos^2 \theta)$~\cite{burset2014transport}.

As $\theta$ denotes the relative orientation between the triplet and singlet components of the pairing potential,
its range can be [$-\pi/2$,$\pi/2$] with respect to the direction of the incoming electron~\cite{burset2014transport}. 
Therefore, the angle-averaged conductance can be obtained after integrating $G(\epsilon,\theta)$ over $\theta$ as,

\bea
&&\tilde{G} (\epsilon)=\int \limits_{-\pi/2}^{\pi/2} G(\epsilon,\theta) \cos \theta ~ d\theta\ .
\label{normG}
\eea

\section{Numerical Results} {\label{sec:IV}}
In this section we present our numerical results for the scattering probabilities, conductance and shot noise. 
Depending on the ratio of triplet to singlet phase of the superconducting pairing potential we consider 
three different regimes of interest: $\Delta_{p}<\Delta_{s}$, $\Delta_{p}=\Delta_{s}$ and $\Delta_{p}>\Delta_{s}$. 
Although we present all our numerical results only for the regime $\Delta_{p}>\Delta_{s}$ which is the interesting regime. 
Also, we show $R_{e \sigma}$, $R_{h \sigma}$, $T_{e \sigma}$ and $T_{h \sigma}$ as functions of 
different parameters of the system only for $\sigma=1$ without loss of generality and hence we use the notation $R_e$, 
$R_h$, $T_e$ and $T_h$ in place of $R_{e\sigma}$, $R_{h\sigma}$, $T_{e\sigma}$ and $T_{h\sigma}$,
respectively throughout our results. Length of the superconducting region and energy of the incident electron are normalized 
by the superconducting coherence length ($\xi$) and amplitude of the pair potential $\Delta_0$, respectively \ie, $L/\xi\rightarrow L$, 
$\epsilon/\Delta_0\rightarrow\epsilon$. Moreover, depending on the doping in the normal metal side we divide our 
study into two categories such as, undoped case where we set $\mu=0$ and finite doping condition 
for which we fix $\mu=5$. Throughout our calculation, the values of some parameters are taken as $Z=2$, $\phi=0$, $e=1$, $h=1$ 
and $U=15$ (for the superconducting region). The chosen value of $U$ makes the superconductor doped and creates 
large Fermi wave-length mismatch between the normal and superconducting regions and also fulfill the requirement of the mean field 
condition of superconductivity \ie\,$\mu+U\gg\Delta_0$. We use the unit where $\Delta_0=1$.

\subsection{Scattering Processes}
In this section we show our numerical results for the scattering probabilities for two different doping conditions. 

\subsubsection{Undoped Regime ($\mu=0$)}
\begin{figure}[!thpb]
\centering
\includegraphics[width=0.95\linewidth]{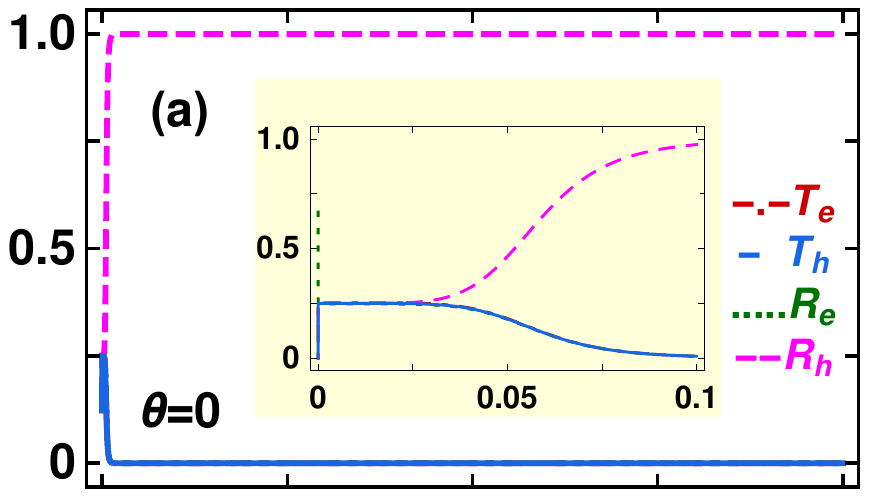}\\
\includegraphics[width=0.96\linewidth]{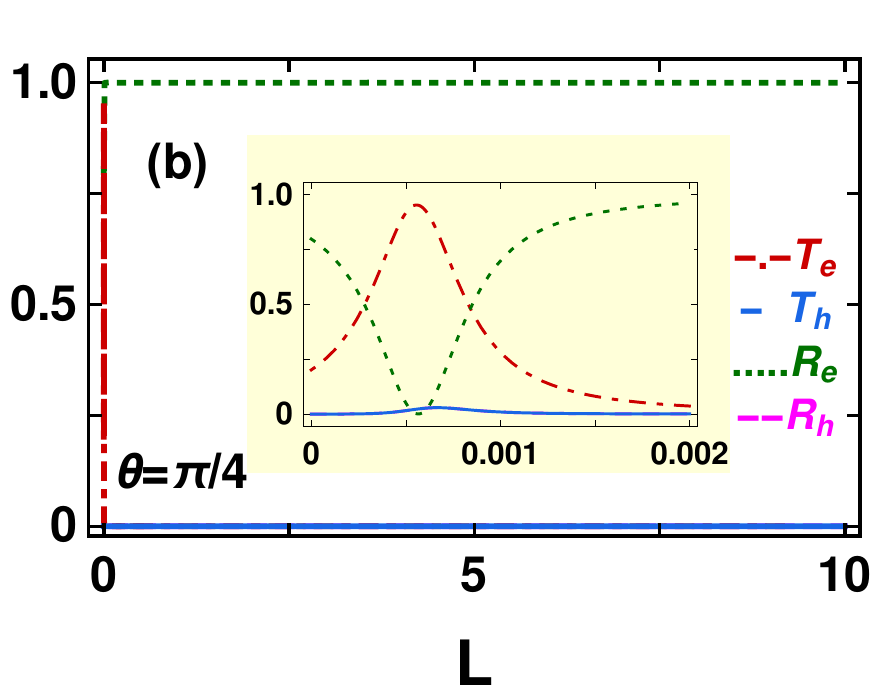}
\caption{(Color online) Quantum-mechanical scattering probabilities (normal reflection $R_e$, Andreev reflection $R_h$, 
elastic co-tunneling $T_e$ and crossed Andreev reflection $T_h$) are plotted as a function of the length ($L$) of 
the superconducting region. In the insets, the behavior of the scattering probabilities are shown when $L\ll \xi$.
Here $\theta=0$ in the upper panel and $\theta=\pi/4$ in the lower panel. 
The value of the other parameters are chosen to be $\mu=0$, $\Delta_{s}=0.25$, $\Delta_{p}=0.75$ and $\epsilon=0$.}
\label{TLmu0}
\end{figure}

In Fig.~\ref{TLmu0} we show all the four possible scattering probabilities $R_e$, $R_h$, $T_e$ and $T_h$ as a function
of the length ($L$) of the superconductor for $\Delta_{p} > \Delta_{s}$ regime, setting incident electron energy $\epsilon=0$. 
With this choice of energy value we are within the superconducting sub-gapped regime. Panel (a) and (b) in Fig.~\ref{TLmu0} 
correspond to $\theta=0$ and $\theta=\pi/4$ respectively. 

It is evident from Fig.~\ref{TLmu0}(a) that for $\theta=0$, AR dominates over all other scattering processes except 
for very small values of $L$. To illustrate this, we show $R_e$, $R_h$, $T_e$ and $T_h$ in the inset of Fig.~\ref{TLmu0}(a),
for small values of $L$ ($L\ll \xi$). Note that all the scattering probabilities are almost identical to each other  
for $L<0.03 \xi$ \ie they occur with almost equal probability of value $\sim 0.25$ which is in sharp contrast to the 
$\Delta_{s}>\Delta_{p}$ regime where $T_h$ (CAR) is vanishingly small. On the other hand, for $L>0.075 \xi$, 
all scattering processes except AR become vanishingly small. 

To investigate whether the above mentioned resonance phenomena persists for other values of $\theta$, we show the 
behavior of $R_e$, $R_h$, $T_e$ and $T_h$ as a function of $L$ in Fig.~\ref{TLmu0}(b) for $\theta=\pi/4$. Note that
the scattering probabilities no longer remain equal to each other for $L\ll \xi$ regime even we set $\Delta_{p} > \Delta_{s}$. 

\begin{figure}[!thpb]
\centering
\includegraphics[width=0.94\linewidth]{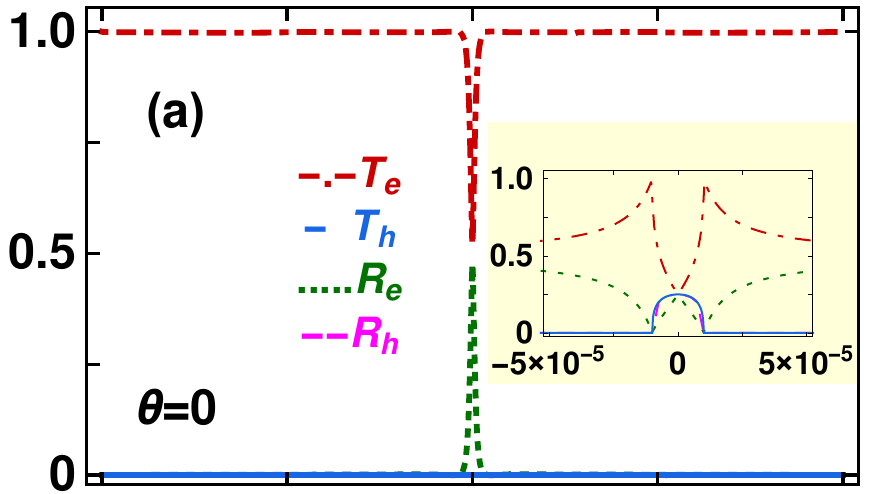}
\includegraphics[width=0.965\linewidth]{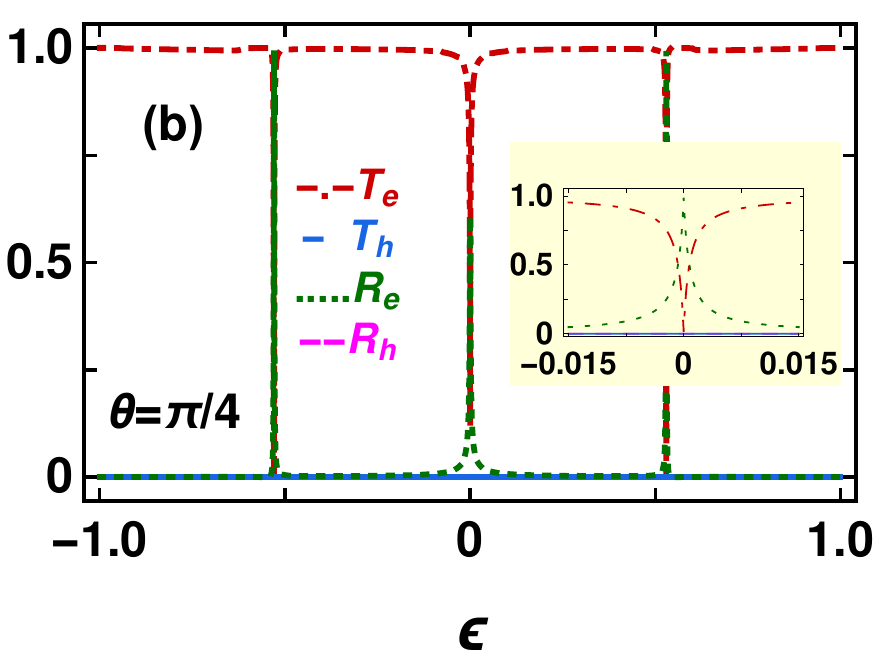}
\caption{ (Color online) The behavior of normal reflection $R_e$, Andreev reflection $R_h$, 
elastic co-tunneling $T_e$ and crossed Andreev reflection $T_h$ are shown as a function of the energy 
($\epsilon$) of the incident electron in the subgapped regime. Here $\theta=0$ in the upper panel and $\theta=\pi/4$ in the lower panel.
We choose $L=0.003 \xi$ and the value of the other parameters are same as in Fig.~\ref{TLmu0}.}
\label{TEmu0}
\end{figure}

Instead, probability for CT dominates over the others and attains the maximum value $\sim 1$ for $L \ll \xi$
which is illustrated in the inset of Fig.~\ref{TLmu0}(b). Nevertheless, as soon as $L$ becomes larger than $\xi$
normal reflection begins to dominate over CT as visible from Fig.~\ref{TLmu0}(b). For $L\gg \xi$ all processes 
except normal reflection die away and the junction becomes perfectly reflecting. Comparing the two cases we can say 
that for $\theta=0$, AR dominates over the other processes when $L\gg \xi$. On the other hand, the contribution for 
normal reflection process becomes dominant for $\theta=\pi/4$ and $L\gg \xi$. For both the $\theta$ values, the 
contribution for the two non-local processes $T_e$ (CT) and $T_h$ (CAR) becomes vanishingly small when $L > \xi$.  

We also analyse the dependence of this resonance phenomenon on the incident electron energy and show 
the corresponding behavior of $R_e$, $R_h$, $T_e$ and $T_h$ as a function of $\epsilon$ in Fig.~\ref{TEmu0}. 
It is evident from the inset of Fig.~\ref{TEmu0}(a) that all the scattering probabilities become equal 
in magnitude ($\sim 0.25$) at $\epsilon=0$. Similar $1/4$ resonance behavior at finite energy had been 
predicted earlier in a superconducting double barrier (NSNSN) structure~\cite{KunduRaoSaha} where the 
superconductor was considered to be a purely singlet one. However, zero energy peak (ZEP) for CAR with peak height 
of $\sim$ 0.25 for a NSN geometry in the $\Delta_{p} > \Delta_{s}$ regime when $\theta=0$ is one of the main results of our paper. 
The physical reason behind the emergence of this ZEP can be attributed to the vanishing of the effective pairing gap 
for $\theta=0$ when $\Delta_{p} = \Delta_{s}$ and changing sign depending on whether $\Delta_{p} > \Delta_{s}$ or $\Delta_{p} < \Delta_{s}$ 
(see Eq.(\ref{delta})). This leads to the appearence of a zero-energy Andreev bound state and zero energy resonance phenomena therein.

\begin{figure}[ht]
{\centering \includegraphics[width=0.95\linewidth]{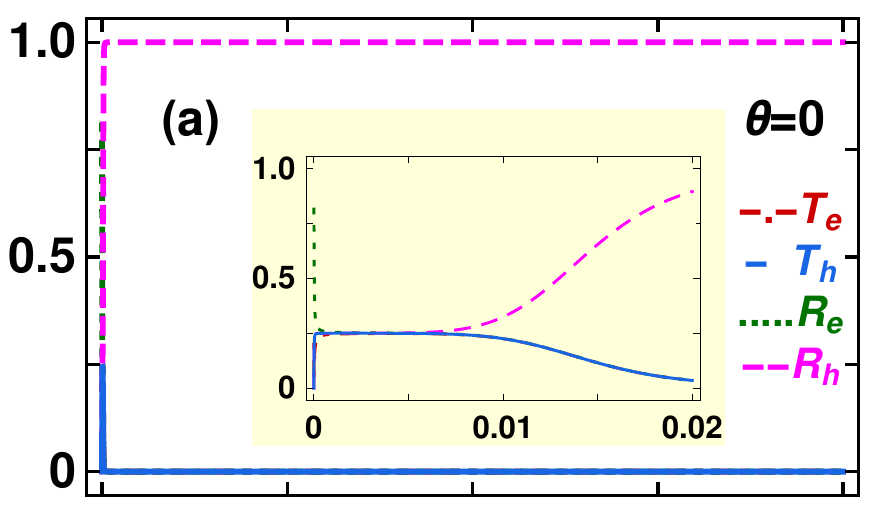}}
{\centering \includegraphics[width=0.95\linewidth]{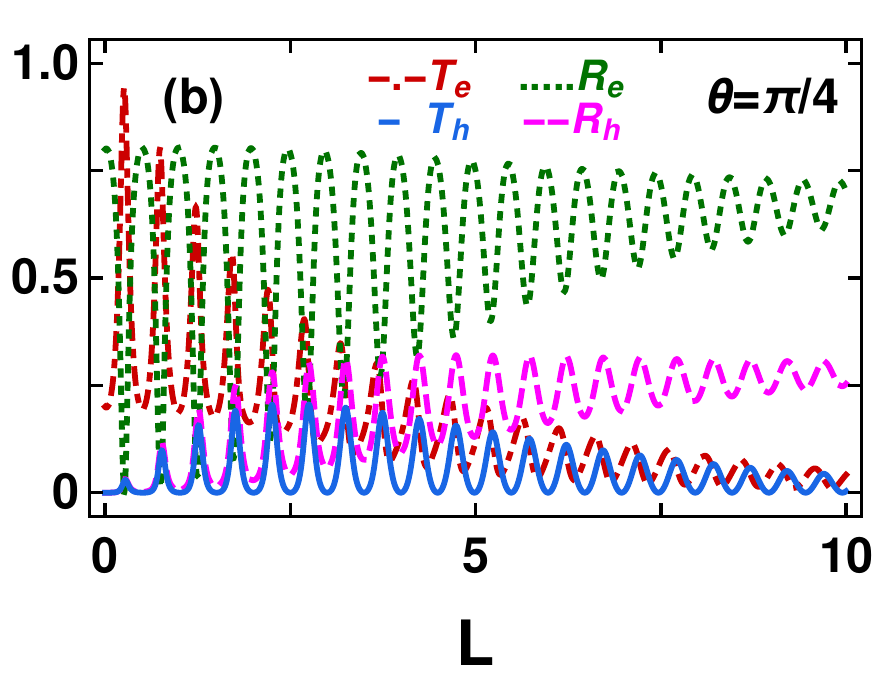}}
\caption{(Color online) The behavior of quantum-mechanical scattering probabilities (normal reflection $R_e$, 
Andreev reflection $R_h$, elastic co-tunneling $T_e$ and crossed Andreev reflection $T_h$) are plotted as a function 
of the length ($L$) of the superconducting region for the doped regime.  
Here we have chosen $\mu=5$, $\Delta_{s}=0.25$, $\Delta_{p}=0.75$ and $\epsilon=0$ with $\theta=0$ (upper panel) 
and $\pi/4$ (lower panel). In the inset of the upper panel, the behavior of $R_e$, $R_h$, $T_e$ and $T_h$ is 
elaborated when $L\ll \xi$.}
\label{TLmu5}
\end{figure}
On the other hand, the behavior of $R_e$, $R_h$, $T_e$ and $T_h$ for $\theta=\pi/4$ is depicted in Fig.~\ref{TEmu0}(b).
We observe that normal reflection has sharp zero energy as well as finite energy peaks, while the CT process has sharp
dips at those energy values. These peaks are clearly shown in the inset of Fig.~\ref{TEmu0}(b).
In this parameter regime AR and CAR probability are always vanishingly small. Note that, for this $\theta$
value, the energy dispersion changes according to Eq.(\ref{delta}) leading to different resonance behavior.

Note that the amplitudes for different scattering processes depend on whether $\mu>\epsilon$ or $\mu<\epsilon$.
This can be understood qualitatively from Eq.~(\ref{keh}). Whether $k_h$ is real or imaginary, it completely depends on the 
relative strength of gate voltage and applied bias which in turn changes the scattering amplitudes. For $\epsilon>\mu$, 
particles can only tunnel through the superconductor resulting in $T_e=1$ as shown in Fig.~\ref{TEmu0}(a) for $\theta=0$.

The zero energy resonance phenomena also survives with the enhancement of the barrier strength $Z$. The reason behind such
resonance structure and ZEP for the AR and CAR can be attributed to the formation of Andreev bound states (ABS) inside the 
proximity induced superconducting region of the NW. The nature of the ABS from the shot noise point of view will be
presented at a later subsection of this article. 

\subsubsection{Doped Regime ($\mu=5$)}
Here, we examine the behavior of the scattering probabilities with the change of doping in the normal metal
while choosing the value of the other parameters same as in the undoped case. Fig.~\ref{TLmu5} depicts the 
variation of $T_e$, $T_h$, $R_e$ and $R_h$ as a function of the length of the superconductor where panel (a) and (b) 
correspond to $\theta=0$ and $\pi/4$, respectively. In Fig.~\ref{TLmu5}(a) we observe that the behavior
of the scattering probabilities qualitatively remains similar to that of the undoped ($\mu=0$) case. 
Here also AR dominates over all other scattering processes and also the $1/4$ resonance phenomena 
takes place below a critical value of $L/\xi$. The latter is depicted in the inset of Fig.~\ref{TLmu5}(a).
Hence, doping has a very small effect on the scattering phenomena when $\theta=0$ and $\epsilon=0$. 

\begin{figure}[!thpb]
\centering
\includegraphics[width=0.96\linewidth]{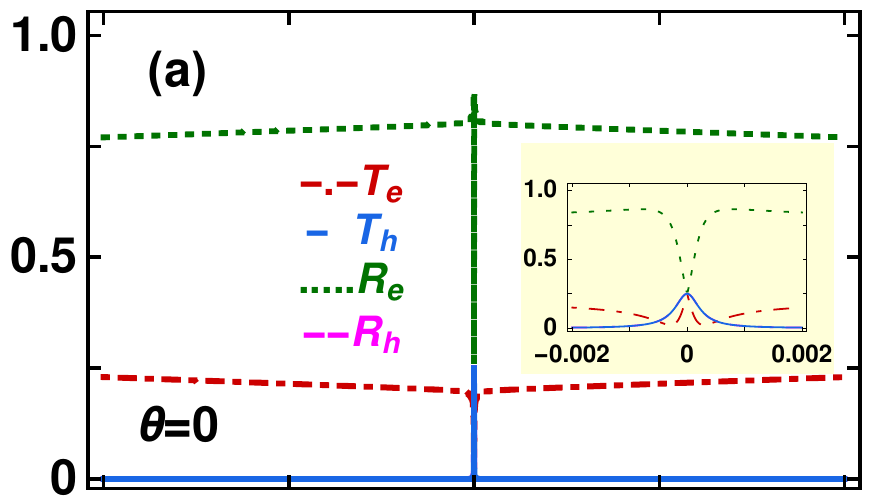} \\
\vspace{-0.11cm}
\includegraphics[width=0.97\linewidth]{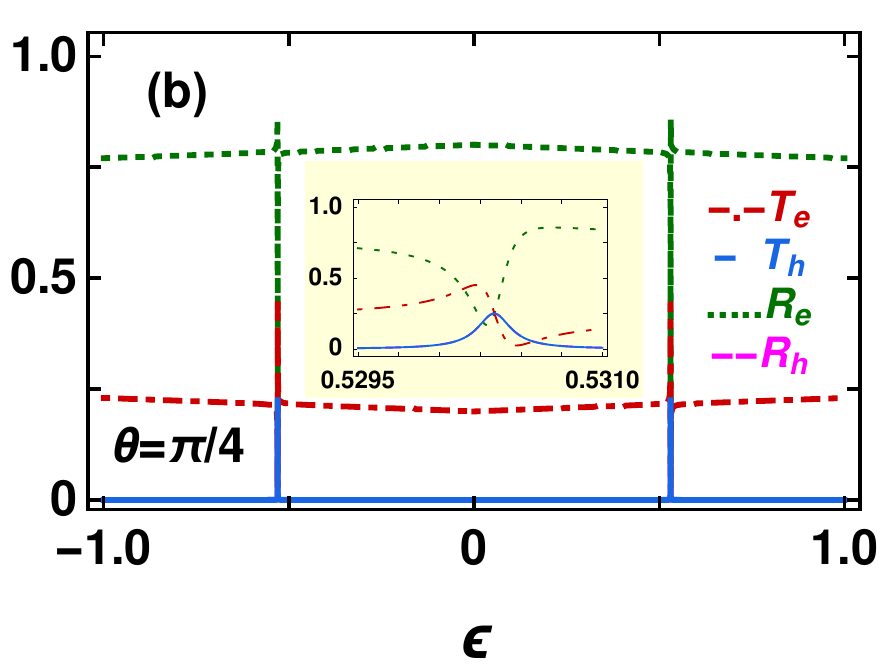}
\caption{(Color online) The behavior of $R_e$, $R_h$, $T_e$ and $T_h$ is shown as a function of energy
of the incident electron in the subgapped regime with $\theta=0$ (upper panel) and $\theta=\pi/4$ (lower panel). 
The value of the other parameters are chosen to be the same as in Fig.~\ref{TLmu5}.}
\label{TEmu5}
\end{figure}

However, if we choose a different value of $\theta$, the effect of doping is much more visible 
in Fig.~\ref{TLmu5}(b) in comparison to Fig.~\ref{TLmu0}(b). All the scattering probabilities 
become oscillatory with respect to $L/\xi$ when we set $\theta=\pi/4$ for finite doped regime. 
The only common feature between the two cases in that with the enhancement of the length of
the superconducting region, normal reflection dominates over all other processes while CT and 
CAR become vanishingly small. These periodic variation with $L$ can be manifested as the 
interference between the electron and hole wave-functions inside the superconducting region. 

In Fig.~\ref{TEmu5} we show the variation of $R_e$, $R_h$, $T_e$ and $T_h$ with incident electron energy
$\epsilon$. Here panel (a) and (b) correspond to $\theta=0$ and $\theta=\pi/4$ while the value of the 
other parameters remain unchanged as in the undoped case. The inset of Fig.~\ref{TEmu5}(a) illustrates 
that AR, CAR and CT processes acquire sharp peak and all of them achieve a value $\sim 0.25$ at zero energy. 
They gradually become vanishingly small for $|\epsilon|> 0.002\Delta_{0}$. On the other hand the probability
for $R_{e}$ becomes close to unity and the junction becomes nearly perfectly reflecting for energy values other 
than zero. On the contrary, for $\theta=\pi/4$, the ZEP no longer exists as depicted in Fig.~\ref{TEmu5}(b). 
There are two resonance points symmetrically situated around $\epsilon \approx \pm 0.5 \Delta_{0}$ in the subgapped regime. 
Both AR and CAR have sharp peaks whereas the other two processes ($R_{e}$ and CT) have dip at those points
(see the inset of Fig.~\ref{TEmu5}(b)). These peaks (dips) are shifted from zero energy due to finite $\theta$ 
in the doped regime. 

Note that all the results presented here is for symmetric barriers placed at the two N-NW interfaces. 
However, our results remain qualitatively unchanged even for asymmetric barrier strengths at the two interfaces.

\subsection{Conductance}
\begin{figure}[!thpb]
\centering
\includegraphics[width=0.98\linewidth]{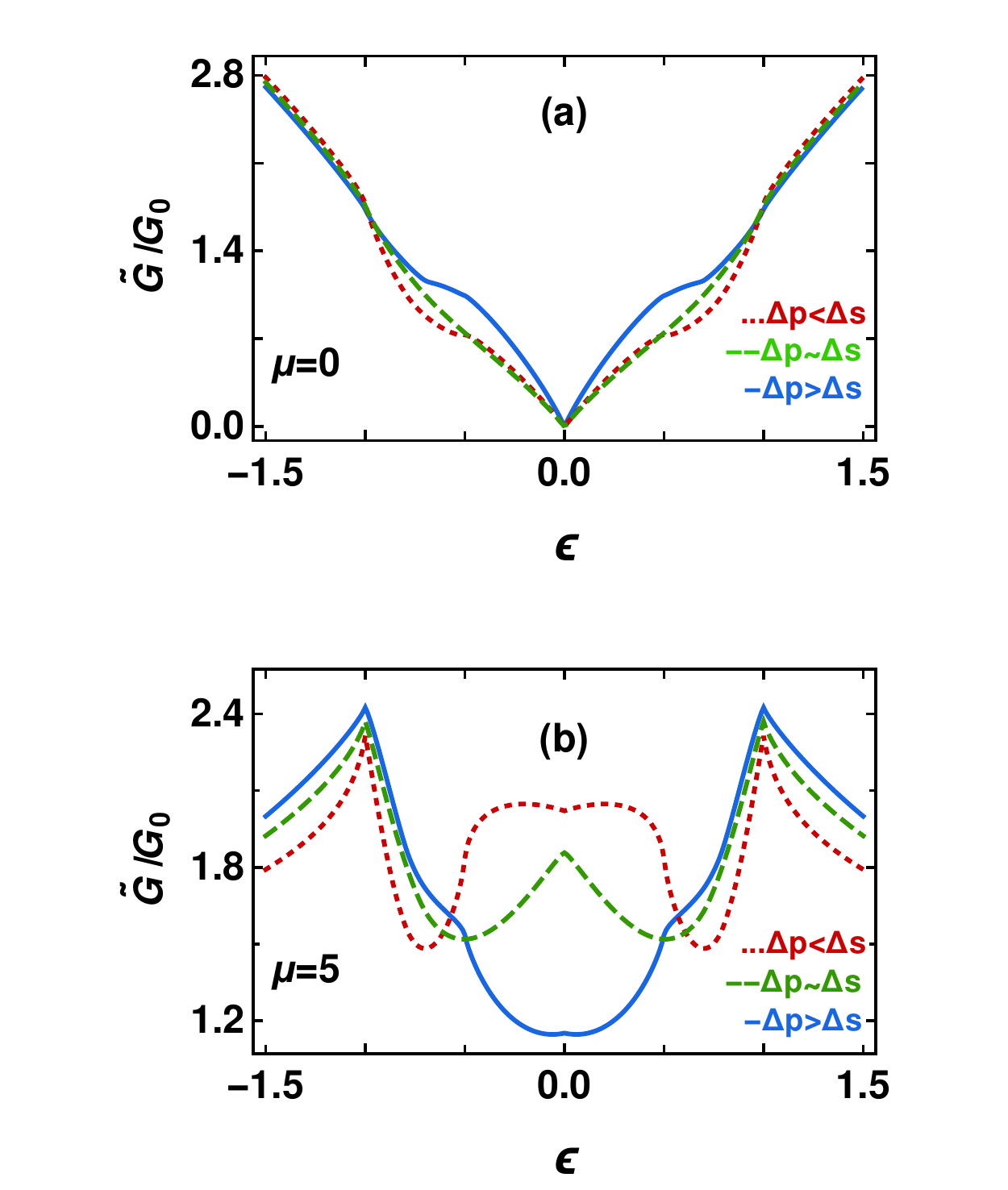}
\caption{(Color online) The behavior of normalized conductance ($\tilde{G}/G_0$) is shown as a function of the 
energy ($\epsilon$) of the incident electron for three different regime of mixed pairing potential
($\Delta_{p} < \Delta_{s}$, $\Delta_{p} \sim \Delta_{s}$ and $\Delta_{p} > \Delta_{s}$). Here we choose
$L=0.75 \xi$. The upper and lower panel correspond to undoped ($\mu=0$) and doped ($\mu=5$) regime.}
\label{GE}
\end{figure}

In this subsection, we study the angle averaged normalized conductance ($\tilde{G}/G_0$) as a function of incident electron energy 
$\epsilon$ using Eq.~(\ref{normG}). The results are presented in Fig.~\ref{GE} where, panel (a) and (b) correspond to 
the undoped ($\mu=0$) and doped ($\mu=5$) case, respectively. Here we have averaged over all possible orientations 
between the singlet and triplet pair potentials. For the undoped case, conductance increases almost linearly with energy 
as shown in Fig.~\ref{GE}(a) irrespective of the ratio of the pairing potential amplitudes. At $\epsilon=0$, averaged
conductance exactly becomes zero for all the three regimes of the pairing potentials. On the contrary, in the doped
regime, conductance behavior is non-monotonic. There are peaks at $\epsilon=\pm \Delta_{0}$ for all the three 
regimes ($\Delta_{p}< \Delta_{s}$, $\Delta_{p}=\Delta_{s}$, $\Delta_{p} > \Delta_{s}$) and these peaks correspond 
to the density of states at the two boundaries of the superconducting gap. In the scattering probability profiles 
we obtain ZEP for CAR and CT processes in the regime $\Delta_{p}> \Delta_{s}$. However in the conductance
profile, we obtain a ZEP when $\Delta_p = \Delta_s$ for the finite doping condition. There is only finite average conductance
(no ZEP) for the other two regimes \ie $\Delta_{p} < \Delta_{s}$ or $\Delta_{p} > \Delta_{s}$ (see Fig.~\ref{GE}(b))

The absence of the ZEP in the orientation averaged conductance profile corresponding to $\Delta_{p} > \Delta_{s}$ 
regime can be explained as follows. At $\epsilon=0$, we have zero conductance corresponding to the regime $\Delta_{p}>\Delta_{s}$ 
for the undoped case. This happens because at $\epsilon=0$ either both CT and CAR probabilities have the same magnitude 
describing the resonance condition or they are vanishingly small as depicted in Fig.~\ref{TEmu0}[(a)-(b)]. 
Taking into account contributions due to all possible orientations ($\theta$) between the singlet and triplet pairings
we finally obtain zero conductance for the undoped case. Although for non-zero $\epsilon$, the contribution in the conductance 
can arise due to entirely CT ($T_{e}$) process whose probability is one for $\epsilon \neq 0$. On the other hand, for the 
doped case we have finite conductance at $\epsilon=0$ after averaging over all possible $\theta$s. The reason behind this 
feature can be attributed to the finite $T_{e}$ contribution for $\theta=\pi/4$ (see Fig.~\ref{TEmu5}(b)). 

\subsection{Shot noise}
This sub-section is devoted to explore the shot noise properties mediated through our NSN junction. Our aim is to
investigate the nature of zero energy resonance as mentioned before via the noise. In general shot noise in a
mesoscopic system arises due to the quantization of the electric charge~\cite{blanter2000shot,beenakker2003quantum}. 
Measurement of shot noise can even be utilized to probe the nature of superconducting wavefunction~\cite{WeiChandrasekhar}. 
Here we neglect thermal noise as throughout our calculation we set the temperature to zero.

The correlation function of the current in the two leads labeled by $i$ and $j$, is defined as~\cite{blanter2000shot},
\bea
S_{ij}(t-t')=<\Delta\hat{I}_i(t)\Delta\hat{I}_j(t^{\prime})+\Delta\hat{I}_j(t^{\prime})\Delta\hat{I}_i(t)>
\label{shot}
\eea
in terms of the operator,
\bea
\Delta\hat{I}_i(t)=\hat{I}_i(t)-<\hat{I}_i(t)>.
\eea
After performing Fourier transform Eq.~(\ref{shot}) becomes,
\begin{align}
S_{ij}(\omega)\delta(\omega+\omega^{\prime})=\frac{1}{2\pi}<\Delta\hat{I}_i(\omega)\Delta\hat{I}_j(\omega^{\prime})
+\Delta\hat{I}_j(\omega^{\prime})\Delta\hat{I}_i(\omega)>
\end{align}
with
\bea
\Delta\hat{I}_i(\omega)=\hat{I}_i(\omega)-<\hat{I}_i(\omega)>.
\eea

We find the expression for zero frequency ($\omega=0$) shot noise cross correlation in terms of the scattering 
amplitudes following Ref.~\onlinecite{anantram1996current}. The general expression for current fluctuation in 
two different leads $i$ and $j$ in presence of an external bias is given by~\cite{anantram1996current},
\bea
S_{ij}&=&\frac{2e^2}{h}\sum_{k,l\in N,S,\alpha,\beta,\gamma,\delta\in e,h} \text{sgn}(\alpha) 
\text{sgn}(\beta) \non \\
&& \int dE A_{k\gamma,l\delta}(i\alpha,E) A_{l\delta,k\gamma}(j\beta,E) f_{k\gamma}(E)[1-f_{l\delta}(E)] \non \\
\label{sij}
\eea
where $A_{k\gamma,l\delta}(i\alpha,E)=\delta_{ik}\delta_{il}\delta_{\alpha\gamma}\delta_{\alpha\delta}
-s_{ik}^{\alpha\gamma\dagger}(E)s_{il}^{\alpha\delta}(E)$.
Here sgn($\alpha$)$=+1$ corresponds to $\alpha=e$ (electron) and sgn($\alpha$)$=-1$ refers to $\alpha=h$ (hole). 
$s_{ik}^{\alpha\gamma}$ represents the scattering amplitude for a particle of type $\gamma$ incident from lead $k$ 
being transmitted to lead $i$ as a particle of type $\alpha$ ($\alpha,\gamma\in e,h$). Eq.~(\ref{sij}) is valid for 
current fluctuations in mesoscopic hybrid junctions when the superconductor region is maintained at a fixed potential~\cite{anantram1996current}. 
Also, we consider zero frequency limit to neglect capacitive component in order to avoid displacement currents due to charging.

It is well known that zero-frequency current fluctuation between two different normal metal leads is always negative 
for fermions~\cite{blanter2000shot}. Nevertheless in presence of a singlet superconductor shot noise cross-correlation 
can be positive depending on the parameter values~\cite{thierrymartin1,thierrymartin2,WeiChandrasekhar,AndyDas}. 
The expression for shot noise in terms of transmission and reflection co-efficients are given in the Appendix~\ref{apndx}. 

In Fig.~\ref{Shot0}[(a)-(b)] we show the behavior of shot noise cross correlation ($S_{ij}$) as a function of the incident
electron energy $\epsilon$ in the regime $\Delta_{p}>\Delta_{s}$ for the undoped ($\mu=0$) case. Here panel $(a)$
and $(b)$ correspond to $\theta=0$ and $\theta=\pi/4$ respectively. We observe that $S_{ij}$ gradually reduces to
-1 for very small range around $\epsilon=0$, but sharply reverts back to zero exactly at $\epsilon=0$ as illustrated
by the inset of Fig.~\ref{Shot0}(a). This sign change of $S_{ij}$ from -1 to 0 reflects the presence of the zero
energy resonance where all the scattering probabilities have equal magnitude of 1/4. Moreover, for energy values
other than zero below the subgapped regime, $S_{ij}$ is exactly zero since $T_{e}=1$ and $R_{e}=R_{h}=T_{h}=0$ 
(see Fig.~\ref{TEmu0}(a)). 
\begin{figure}[!thpb]
\centering
\includegraphics[width=0.95\linewidth]{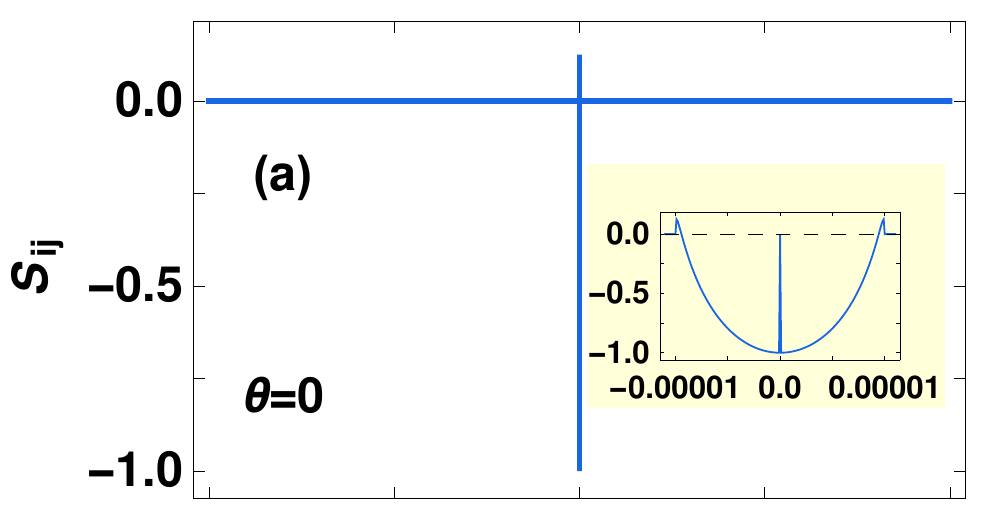}
\includegraphics[width=0.98\linewidth]{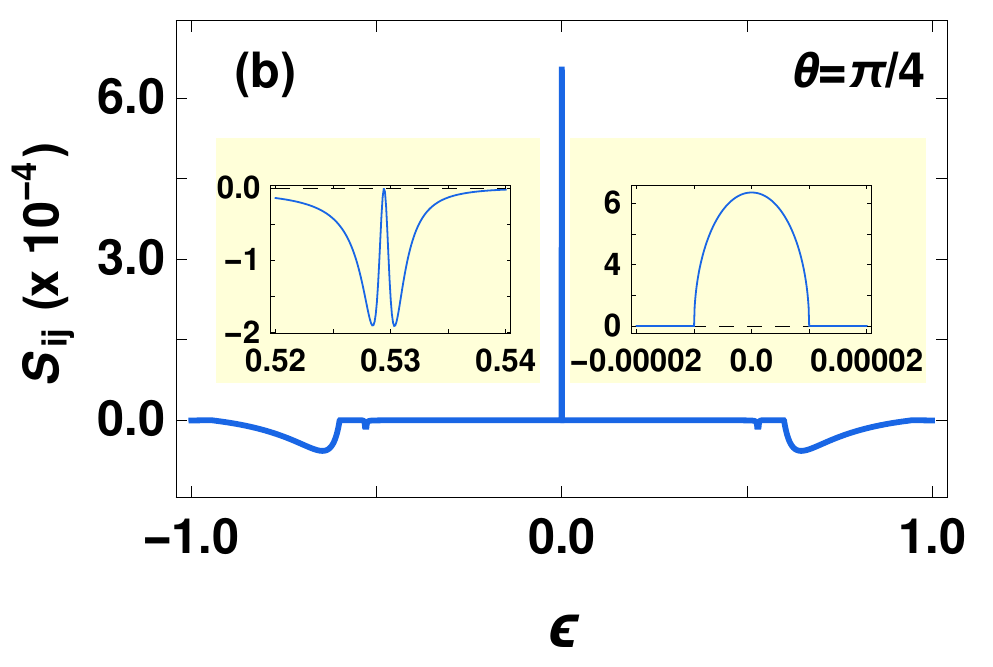}
\caption{(Color online) The behavior of shot noise cross-correlation ($S_{ij}$) is shown as a function of energy 
($\epsilon$) of the incident electron. Here $L=0.003 \xi$, $\Delta_{s}=0.25$, $\Delta_{p}=0.75$ and $\theta=0$ (upper panel), 
$\theta=\pi/4$ (lower panel) with $\mu=0$.}
\label{Shot0}
\end{figure}
On the other hand, there are sharp positive peaks in the $S_{ij}$ profile for $\theta=\pi/4$, as shown in Fig.~\ref{Shot0}(b). 
For $\epsilon=0$ and $\epsilon\approx 0.5$, $R_{e}=1$ and $R_{h}=T_{e}=T_{h}=0$ as depicted in Fig.~\ref{TEmu0}(b). Also
$T_{e}=1$ for the other values of energy. Hence, shot noise cross correlation $S_{ij}$ is vanishingly small ($\sim \rm 10^{-4}$) 
compared to the $\theta=0$ case. 

\begin{figure}[!thpb]
\centering
\includegraphics[width=0.99\linewidth]{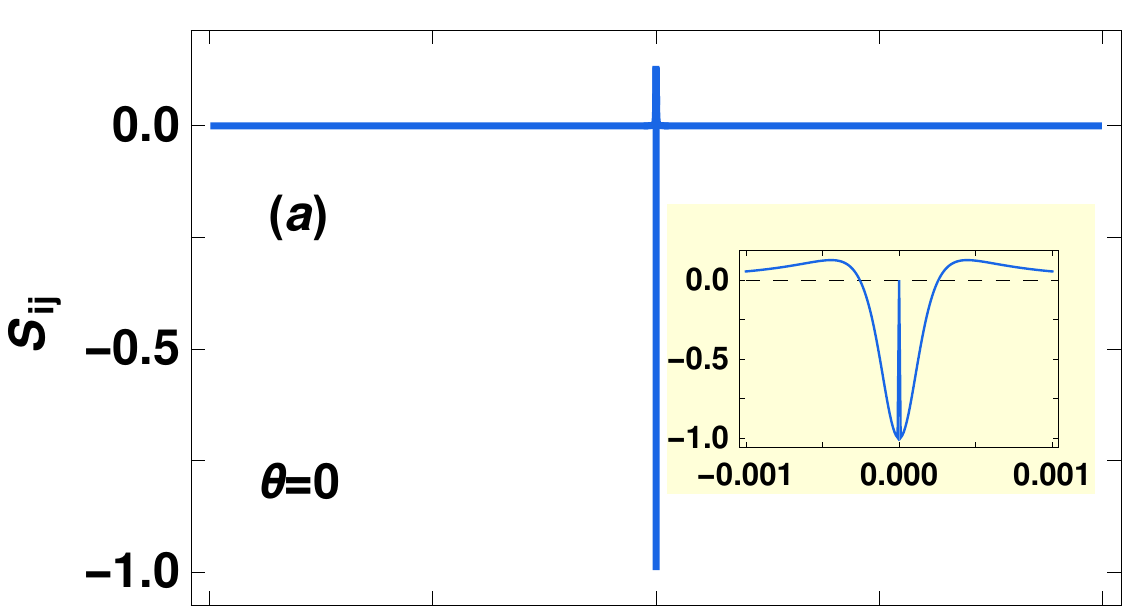}
\includegraphics[width=0.99\linewidth]{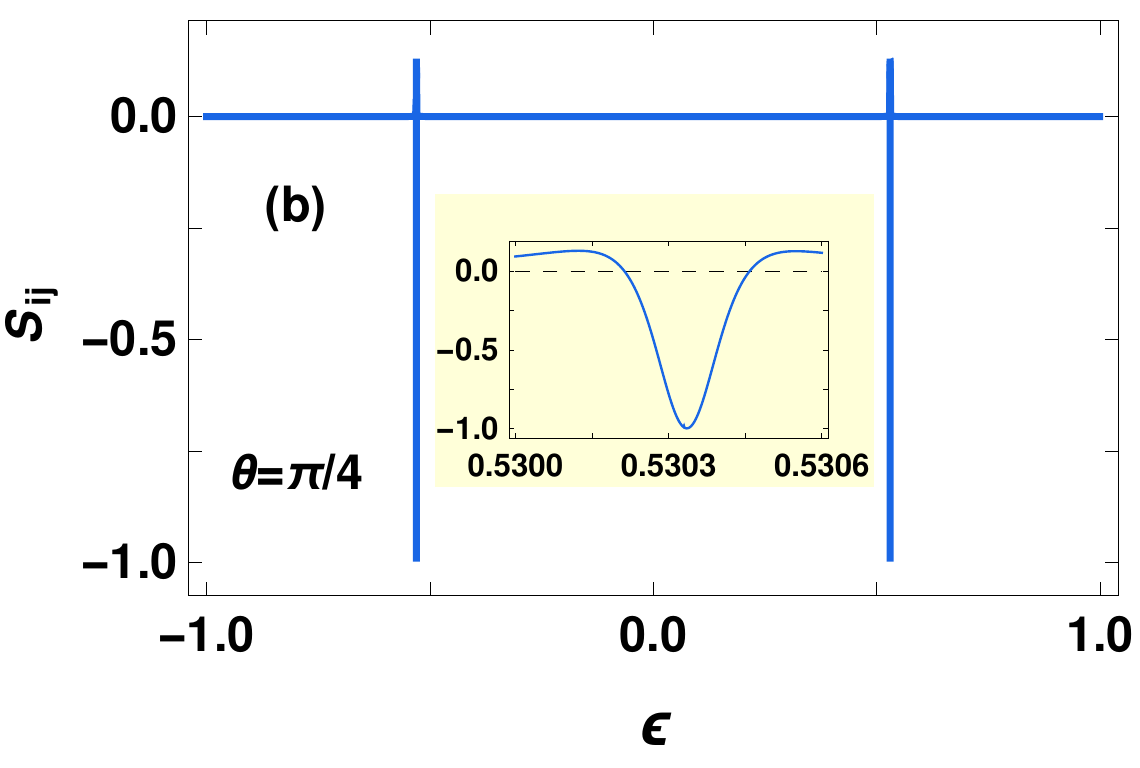}
\caption{(Color online) The feature of shot noise cross-correlation ($S_{ij}$) is depicted  as a function of energy ($\epsilon$) 
of the incident electron. Here $\mu=5$ and $\theta=0$ (upper panel) and $\theta=\pi/4$ (lower panel). We choose the value 
of the other parameters same as in Fig.~\ref{Shot0}.}
\label{Shot5}
\end{figure}

Similar to the undoped case, we also calculate the zero frequency shot noise cross-correlation for the doped system. 
Our results are shown in Fig.~\ref{Shot5}[(a)-(b)]. We qualitatively obtain the similar behavior for $S_{ij}$ as in
the undoped case. As depicted in Fig.~\ref{TEmu5}(a), at $\epsilon=0$ all the scattering probabilities have the 
same value of 0.25 resulting in zero $S_{ij}$ (see Fig.~\ref{Shot5}(a)). On the other hand, for $\theta=\pi/4$,
$S_{ij}$ becomes -1 in a small range of around $\epsilon\approx\pm 0.5\Delta_{0}$. Another interesting feature is that $S_{ij}$ changes 
sign from negative to positive for both the cases $\theta=0$ (around $\epsilon=0$) and $\theta=\pi/4$ (around $\epsilon\approx\pm 0.5\Delta_{0}$) 
as depicted in the insets of Fig.~\ref{Shot5}(a) and Fig.~\ref{Shot5}(b) respectively. This transition of the shot noise cross-correlation 
from negative to positive value is in contrast to that case of purely singlet superconductor where shot noise cross correlation 
is only positive~\cite{thierrymartin1,WeiChandrasekhar,AndyDas}.

So far, the behavior of shot noise cross-correlation has been discussed for particular values of $\theta$ which is
the orientation between the triplet to singlet amplitude of the superconductor. To obtain the angle averaged shot noise 
we integrate over all possible orientations as,
\bea
\tilde{S}_{ij}=\int \limits_{-\pi/2}^{\pi/2} S_{ij} \cos \theta ~d\theta.
\eea

\begin{figure}[!thpb]
\centering
\includegraphics[width=0.99\linewidth]{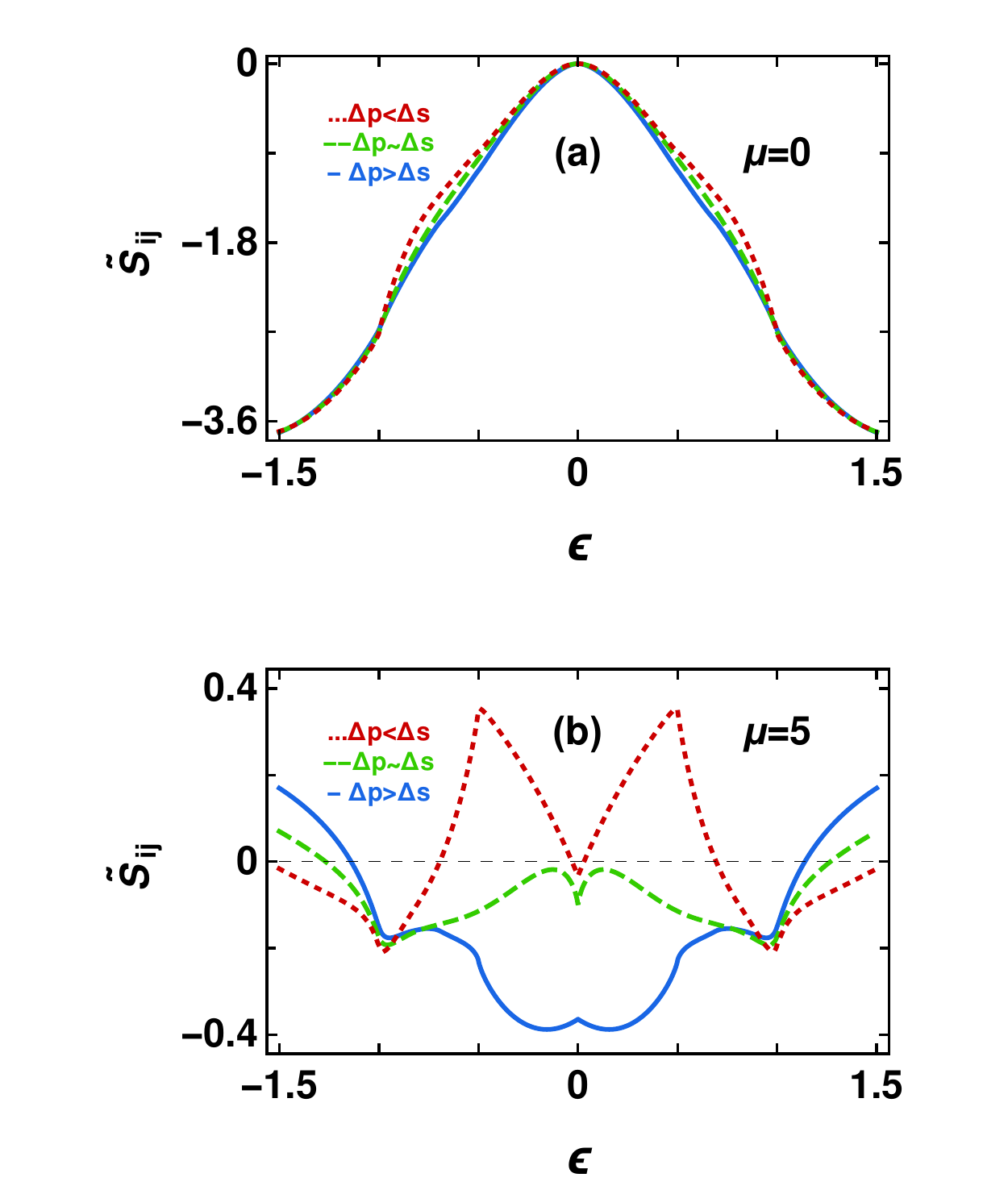}
\caption{(Color online) Angle averaged shot noise cross-correlation ($\tilde{S}_{ij}$) is shown as a function of incident 
electron energy ($\epsilon$) for three different regimes of mixed pairing potential ($\Delta_{p} < \Delta_{s}$, 
$\Delta_{p} \sim \Delta_{s}$ and $\Delta_{p} > \Delta_{s}$). We have chosen $L=0.75 \xi$. The upper and lower panel 
represent $\mu=0$ and $\mu=5$ cases respectively.}
\label{Shotavg}
\end{figure}

The behavior of angle averaged shot noise $\tilde{S}_{ij}$ is presented as a function of the energy of the incident 
electron in Fig.~\ref{Shotavg}. For $\mu=0$, $\tilde{S}_{ij}$ vanishes at $\epsilon=0$ irrespective of the ratio of 
the pairing potential amplitudes. Moreover, the feature of $\tilde{S}_{ij}$ is monotonic similar to the conductance
(see Fig.~\ref{GE}(a)) for $\epsilon\neq0$. 

On the other hand, there are crossovers from positive to negative and vice-versa for the doped case 
corresponding to all the three regimes of the pairing amplitudes. Nevertheless, $\tilde{S}_{ij}$ is always 
negative for $\Delta_{p} \geq \Delta_{s}$ in the subgapped regime \ie $\epsilon \leq \Delta_{0}$. However 
when the $s$ wave pairing amplitude dominates over that of the $p$ wave ($\Delta_{s} > \Delta_{p}$), 
$\tilde{S}_{ij}$ remains positive over the major range of the subgapped regime and changes sign around  
$\epsilon \approx 0.75 \Delta_{0}$. The emergence of negative shot noise cross correlation ($\tilde{S}_{ij}$)
in the $\Delta_{p} \geq \Delta_{s}$ regime in contrast to the positive $\tilde{S}_{ij}$ in the $\Delta_{p} < \Delta_{s}$
regime is another main result of our article. 

\section{Summary and Conclusions} {\label{sec:V}}
To summarize, we have explored the conductance and shot noise phenomena through a NSN junction where the superconductor 
is characterized by a mixture of both the spin-singlet and spin-triplet pairings. Our NSN set up comprises of a 1D NW
placed in close proximity to a bulk superconductor with mixed pairing of singlet and triplet type (\eg~NCS superconductor).
The NW is also coupled to two normal metal leads. Depending on the ratio of their pairing amplitudes and the doping 
concentration in the normal metal region we study the behavior of scattering amplitudes, conductance and zero frequency 
shot noise for three different regimes ($\Delta_{p}>\Delta_{s}$, $\Delta_{p}=\Delta_{s}$, $\Delta_{p}<\Delta_{s}$). 
The main feature we obtain in this geometry is the appearance of a zero energy resonance. At the resonance, probability for 
all the four possible scattering processes (reflection, AR, CT and CAR) acquire same magnitude (1/4) when chiral triplet pairing 
amplitude dominates over the singlet one. The angle averaged conductance also exhibits a zero energy peak in the doped regime.
Moreover, for a chosen orientation ($\theta$) between the singlet and triplet pairing, zero frequency shot noise cross correlation
exhibits positive to negative transition in the chiral triplet pairing dominated regime. However, for the doped regime, 
angle averaged shot noise remains negative in the subgapped regime when $\Delta_{p}\geq\Delta_{s}$ and becomes positive 
in the opposite regime ($\Delta_{s}>\Delta_{p}$). Very recently, transition from positive to negative shot noise cross-correlation 
also has been reported in the context of Majorana bound states~\cite{haim2015signatures,haim2015current,tripathi2016fingerprints}. 

As far as practical realization of our NSN structure is concerned, a NW may be possible to fabricate in close proximity to a 
NCS superconductor for \eg~$\rm Mo_{3}Al_{2}C$, $\rm BiPd$ etc.~\cite{karki2010structure,mintumondal}. Such superconductors posses a 
coherence length $\xi \approx 10-20~\rm nm$ for critical magnetic field $H_{c2}(0) \approx 1.2-1.5~\rm T$ as reported in 
Ref.~\onlinecite{karki2010structure,mintumondal}. Hence, our findings for the angle averaged conductance and shot noise cross-correlation 
may be realizable in a proximity induced NW where the length of the superconducting region can be $L=0.75\xi \sim\ 5-15~\rm nm$. 
Our setup may also be used for making future generation entangler devices with unconventional superconductor~\cite{recher,LesovikMatin,Rakesh}.

\acknowledgments{We acknowledge Arun M. Jayannavar for valuable discussions.}

\appendix*
\section{Expression for the shot noise} \label{apndx}
\setcounter{equation}{0} 
We study the current cross-correlation in our NSN geometry at zero temperature and zero frequency limit following~\cite{anantram1996current} . 
The shot noise contributions arising from different scattering amplitudes can be separated in terms of the normal reflection, AR, CT and CAR
amplitudes as follows.
\bea
S_{ij}^{ee}(\epsilon)&=&-\frac{2e^2}{h}[(t_e(\epsilon)r_e^*(\epsilon)+r_e(\epsilon)t_e^*(\epsilon))^2\non \\ &&
+(t_h(\epsilon)r_{h}^*(\epsilon)+r_h(\epsilon)t_h^*(\epsilon))^2], \non \\
S_{ij}^{eh}(\epsilon)&=&\frac{4e^2}{h}\lvert {r_h(\epsilon)t_e^*(\epsilon)
+t_h(\epsilon)r_e^*(\epsilon)}\rvert^2 .
\label{sijphase}
\eea
Hence the total shot noise reads,
\bea
S_{ij}(\epsilon)&=&S_{ij}^{ee}(\epsilon)+S_{ij}^{eh}(\epsilon)
\eea
where, $S_{ij}^{ee}$, $S_{ij}^{eh}$ represent the cross-correlation corresponding to the phenomenon of CT and CAR respectively.
Also, due to particle-hole symmetry we can write
\bea
S_{ij}^{ee}(\epsilon)&=&S_{ij}^{hh}(\epsilon) \non \\
S_{ij}^{eh}(\epsilon)&=&S_{ij}^{he}(\epsilon).
\eea
Here we scale $r_h$ and $t_h$ by $\sqrt{\frac{k_h}{k_e}}$ (\ie after scaling we have 
$r_h \equiv r_h \sqrt{\frac{k_h}{k_e}}$ and $t_h \equiv t_h \sqrt{\frac{k_h}{k_e}}$) in order to maintain the 
probability conservation (unitarity) as discussed earlier. All these scattering amplitudes are complex and hence 
they can be expressed as follow,
\bea
r_h=\lvert r_h \rvert e^{i \theta_1}, ~~~ r_e=\lvert r_e \rvert e^{i \theta_2} \non\\
t_h=\lvert t_h \rvert e^{i \phi_1}, ~~~ t_e=\lvert t_e \rvert e^{i \phi_2}\non
\eea
where $\theta_1$, $\theta_2$, $\phi_1$, $\phi_2$ are the phase factors of the corresponding complex scattering amplitudes. 
They play crucial role in determining the nature of the shot noise which can be realized from Eq.~(\ref{sijphase}). 
We emphasize the scenario where all the scattering probabilities are equal in magnitude (0.25) \ie the zero energy resonance condition. 
As mentioned earlier, we obtain this zero energy resonance for both undoped ($\mu=0$) and doped ($\mu\neq0$) condition. 
This, at zero energy, the expression for shot noise cross-correlation takes the form,
\bea
S_{ij}(\epsilon)&=&-\frac{e^2}{2h}[\cos^2(\phi_2-\theta_2)+\cos^2(\phi_1-\theta_1)]\non\\
&&+\frac{e^2}{2h}[1+\cos(\theta_1+\theta_2-\phi_1-\phi_2)]\ .
\label{shotreso}
\eea
From Eq.~(\ref{shotreso}) it is evident that shot-noise correlation depends only on the phases of different 
scattering amplitudes at resonance. If the phases cancel out among each other then $S_{ij}$ becomes zero
which we obtain at the resonance.

\bibliography{bibfile}{}
-----------------------------

 \end{document}